\documentclass[10pt,a4paper,openright,tablecaptionabove]{scrartcl}
\usepackage{subfig}
\usepackage[nochapters, pdfspacing]{classicthesis}
\usepackage[english]{babel}
\usepackage{bbm}
\usepackage{url}
\usepackage{path}
\usepackage{amssymb}
\usepackage{amsmath}
\usepackage{amsthm}
\usepackage{cancel}
\usepackage{indentfirst}
\usepackage[pdftex]{graphicx}
\usepackage{eucal}
\usepackage{appendix}
\usepackage[utf8]{inputenc}
\usepackage{geometry}
\usepackage{arsclassica}
\usepackage{verbatim}
\usepackage{nameref}
\usepackage{alltt}
\usepackage{pgfplots}

\usepackage[font=small,labelfont=bf]{caption} 

\newcommand{\figref}[1]{\figurename~\ref{#1}}

\graphicspath{{images/}} % cartella dove sono riposte le immagini
\areaset[current]{480pt}{680pt}
\begin{document}

%=========Abstract================
\thispagestyle{empty}
\begin{center}
\Large{{\bf The classical Taub-Nut System: factorization, Spectrum Generating Algebra and solution to the equations of motion}}
\end{center}
\vskip 0.5cm
\begin{center}
\textsc{Danilo Latini}\footnote{E-mail: latini@fis.uniroma3.it}, \textsc{Orlando Ragnisco}\footnote{E-mail: ragnisco@fis.uniroma3.it}
\end{center}
\vskip 0.5cm
\begin{center}
\small{Department of Mathematics and Physics and INFN, Roma Tre University, Via della Vasca Navale 84, I-00146 Rome, Italy}
\end{center}

\vskip  2cm
\begin{center}
Abstract
\end{center}

\noindent $\mathcal{T}$he formalism of SUSYQM ({SUperSYmmetric Quantum Mechanics}) is properly modified in such a way to be suitable for the description and the solution of a classical maximally superintegrable Hamiltonian System, the so-called Taub-Nut system, associated with the Hamiltonian:
$$ \mathcal{H}_\eta ({\mathbf{q}}, {\mathbf{p}}) = \mathcal{T}_\eta ({\mathbf{q}}, {\mathbf{p}}) + \mathcal{U}_\eta({\mathbf{q}}) = \frac{|{\mathbf{q}}| {\mathbf{p}}^2}{2m(\eta + |{\mathbf{q}}|)} - \frac{k}{\eta + |{\mathbf{q}}|} \quad  (k>0, \eta>0) \, .$$
\noindent In full agreement with the results recently derived by A. Ballesteros et al. for the quantum case, we show that  the classical Taub-Nut system shares a number of essential features with the Kepler system, that is just its Euclidean version arising  in the limit $\eta \to 0$,  and for which  a   ``SUSYQM" approach has been recently introduced by S. Kuru and J. Negro. 
In particular, for positive $\eta$ and negative energy the motion is always periodic; it turns out that the period depends upon $ \eta$ and goes to the Euclidean value as $\eta \to 0$. Moreover,  the maximal superintegrability  is preserved  by the $\eta$-deformation, due to the existence of a larger symmetry group  related to   an $\eta$-deformed Runge-Lenz vector, which ensures that  in $\mathbb{R}^3$ closed orbits are again ellipses.  In this context, a deformed version of the third Kepler's law is also recovered.
The closing section is devoted to a discussion of the $\eta<0$ case, where new and partly unexpected features arise.

\section{Introduction}

\noindent We consider the classical Hamiltonian in $\mathbb{R}^N$ given by:
\begin{equation}
\label{hclassica}
\mathcal{H}_\eta ({\mathbf{q}}, {\mathbf{p}}) = \mathcal{T}_\eta ({\mathbf{q}}, {\mathbf{p}}) + \mathcal{U}_\eta({\mathbf{q}}) = \frac{|{\mathbf{q}}| {\mathbf{p}}^2}{2m(\eta + |{\mathbf{q}}|)} - \frac{k}{\eta + |{\mathbf{q}}|} ,
\end{equation}
where $k$ and  $\eta$ are real parameters, ${\mathbf{q}}=(q_1,\dots,q_N)$, ${\mathbf{p}}=(p_1,\dots,p_N)$ $\in \mathbb{R}^N$ are conjugate coordinates and momenta, and ${\mathbf{q}}^2\equiv  |{\mathbf{q}}|^2=\sum_{i=1}^Nq_i^2$. We recall that   $\mathcal{H}_\eta$ has been proven to be a maximally superintegrable Hamiltonian by making use of symmetry techniques~\cite{sigma7011}. This means that $\mathcal{H}_\eta $ is endowed with the maximum possible number $(2N-1)$ of  functionally independent constants of motion (including $\mathcal{H}_\eta $ itself). In fact, besides the integrals of motion provided by the $\mathfrak{so}(N)$ symmetry, $\mathcal{H}_\eta $ is endowed with an $\eta-$deformed $N$D Laplace--Runge--Lenz vector ${\mathbf R}$ implying the existence of $N$ additional constants of motion coming from the components of ${\mathbf R}$, which are given by:
\begin{equation}
\mathcal{R}_i = \frac{1}{m}\sum_{j=1}^N p_j (q_j p_i - q_i p_j) + \frac{q_i}{|{\mathbf{q}}|} (\eta \mathcal{H}_\eta + k),\qquad i=1,\dots, N.
\label{xb}
\end{equation}
The squared modulus of  ${\mathbf R}$ is radially symmetric, and turns out to be expressible in terms of $\mathcal{H}_\eta$ 
and ${\mathbf L}^2$:
\begin{equation}
{\mathbf R}^2=  \sum_{i = 1}^N \mathcal{R}_i^2 = \frac{2 \mathbf L^2}{m} \mathcal{H}_\eta + (\eta \mathcal{H}_\eta + k)^2 .
\label{Rungelenz}
\end{equation}
As a matter of fact the system associated with $\mathcal{H}_\eta$ ({\ref{hclassica}) under the canonical symplectic structure can be considered as a genuine (maximally superintegrable) $\eta$-deformation of the $N$D usual Kepler-Coulomb (KC) system, since the limit $\eta \rightarrow 0$ of $\mathcal{H}_\eta$ (\ref{hclassica})  yields:
\begin{equation}
\mathcal{H}_0 = \frac{\mathbf p^2}{2m} - \frac{k}{|{\mathbf{q}}|} .
\label{eq:ru}
\end{equation}
Moreover  $\mathcal{H}_\eta$   can be  naturally related to the Taub-NUT system~\cite{Ma82,AH85,GR86,FH87,GR88,IK94,IK95,uwano,BCJ,BCJM,GW07,JL}  since   $\mathcal{M}^N$ can be regarded as  the (Riemannian)   $N$D Taub-NUT space~\cite{annals324}. 
It is also known that according to the Perlick classification \cite{Perlick, Ann2009,Bertrand2,CMP} the system \eqref{hclassica} pertains to the class II, and thus it has to be regarded as an ``intrinsic oscillator".  In the sequel  it will be clear that with respect to the Euclidean KC it plays an analogous role to the Darboux III  (D-III in the following)  in comparison with the standard harmonic oscillator \cite{PD}.

Actually it turns out ??mutatis mutandis", the solution to the classical equations of motion for D-III  can be found on the same footing, through the factorisation of the corresponding classical Hamiltonian. The results concerning D-III will be published later in a larger more general paper \cite{future},  where the quantum cases will be also investigated, mostly through the \emph{Shape Invariant Potentials} approach.

\section{Classical Taub-Nut: factorization,  Spectrum Generating Algebra and solution to the equations of motion}

\noindent 
In the following we drastically simplify our setting, and  limit our considerations to the \emph{physical} (i.e. $3$-dimensional) case. As mentioned in the abstract we adapt and   in a sense generalise the construction and the results derived in \cite{Kuru}.

\noindent We will study the Hamiltonian 
\begin{equation} 
H=T(r,p)+V_{eff}(r)=\frac{r p^2}{2m(r+\eta)}+\frac{ l^2}{2mr(r+\eta)}-\frac{k}{r+\eta} = \mathcal{K}(r) H_0 \,,
\label{hamm}
\end{equation}
\noindent 
where $m$, $k$ and $l$ are positive constants, $\eta$ is the deformation parameter, $p \equiv p_r$ is the radial momentum, $H_0$ is the ``undeformed''  Kepler-Coulomb Hamiltonian and $\mathcal{K}(r) \doteq \frac{r}{r+\eta}$. In (\ref{hamm}) we introduced the radial coordinate $r:=|\mathbf{q}|$, canonically conjugated to $p$.

\noindent
The main idea is to use the framework of SUSYQM in the context of classical mechanics to derive algebraically the classical trajectories (see Ref.$\,$\cite{KuruNegro}). Let us then consider the  Hamiltonian \eqref{hamm} written in a slightly different form, namely:
\begin{equation}
H=\frac{r}{r+\eta} \biggl(\frac {p^2}{2m}+ \frac{l^2}{2mr^2}-\frac{k}{r} \biggl) \, .
\label{ham}
\end{equation}
\noindent
Multiplying both sides of \eqref{ham} by $r(r+\eta)$ we get:
\begin{equation}
r(r+\eta)H=r^2\biggl(\frac{ p^2}{2m}+\frac{l^2}{2mr^2}-\frac{k}{r}\biggl)=\frac{1}{2m}(r^2p^2+l^2-2 m k r) .
\label{fact}
\end{equation}
\noindent Now, as it has been done  in the undeformed case by Kuru and Negro (\cite{KuruNegro}), at any $r$ we can  factorize \eqref{fact} as follows:
\begin{equation}
r^2p^2-2 m r(k+\eta H)-2 m r^2H=A^+A^-+\gamma(H)=-l^2 \, ,
\label{factor}
\end{equation}
\noindent where for the time being $A^+$, $A^-$ are unknown functions of  $r$, $p$. Paraphrasing \cite{Kuru} we make the following \emph{ansatz} for $A^+$, $A^-$
\begin{equation}
A^{\pm} = \biggl(\mp i r p + a r \sqrt{-H}  +\frac{b(H)}{\sqrt{-H}} \biggl)e^{\pm f(r,p)} \, .
\label{eq:ansatz}
 \end{equation}
\noindent The ``arbitrary function'' $f(r,p)$ will be determined by requiring the closure of the Poisson algebra generated
by $H$ and $A^{\pm}$. More precisely, we impose:
\begin{align}
& \{H,A^{\pm} \}=\mp i \alpha(H) A^{\pm} \\
&\{A^+,A^-\}= i \beta(H) \, ,
\label{eq:poisson}
\end{align}
\noindent where  the functions $\alpha$, $\beta$ wait to be determined. Inserting $A^{\pm}$ in \eqref{factor} we get
\begin{equation}
a=\sqrt{2 m} \, , \qquad b(H)=-\sqrt{\frac{m}{2}}(k+\eta H) \, , \qquad \gamma(H)=\frac{m (k+\eta H)^2}{2 H} \, ,
\label{constants}
\end{equation}
\noindent
and requiring that  $A^{\pm}$ obey the proper Poisson brackets we arrive at
\begin{equation}
f(r,p)=-i \sqrt{\frac{2}{m}}\frac{r p \sqrt{- H}}{(k-\eta H)}\,\, , \quad \alpha(H) =-\sqrt{\frac{2}{m}}\,\frac{2 H \sqrt{-H}}{(k-\eta H)} \,\, , \quad \beta(H)=\sqrt{2m}\,\frac{(k+ \eta H)}{\sqrt{-H}} \, ,
\label{poissonAA}
\end{equation}
\noindent and finally: 
\begin{equation}
A^{\pm} = \biggl(\mp i r p +  r \sqrt{- 2 m H}  -\sqrt{\frac{m}{2}}\frac{(k+\eta H)}{\sqrt{-H}} \biggl)e^{\mp i \sqrt{\frac{2}{m}}\frac{r p \sqrt{- H}}{(k-\eta H)}}
\label{ladder}
\end{equation}
\begin{equation}
\{H,A^{\pm} \}=\pm i \sqrt{\frac{2}{m}}\,\frac{2 H \sqrt{-H}}{(k-\eta H)} A^{\pm} \,\, , \qquad  \{A^+,A^- \}= i \sqrt{2m}\,\frac{(k+ \eta H)}{\sqrt{-H}} \, .
\label{poissonbrackets}
\end{equation}
 \noindent A mandatory requirement is that in the limit $\eta \to 0$  one gets back the undeformed Poisson algebra that is in fact, for $2m=k=1$, the result found in \cite{Kuru}.
\noindent
To make the identification even more perspicuous we can introduce $A_0 \doteq \sqrt{\frac{m}{2}}\frac{(k+ \eta H)}{\sqrt{-H}}$ entailing the following $\mathfrak{su}(1,1)$ algebra relations:
\begin{equation}
\{A_0,A^{\pm} \}=\mp i A^{\pm} \,\, , \quad \{A^+,A^- \}=  2 i A_0\, .
\label{sualgebra}
\end{equation}
\noindent Now we can define the ``time-dependent constants of the motion''
\begin{equation}
Q^{\pm}=A^{\pm} e^{\mp i \alpha(H) t} \, ,
\label{eq:constant}
\end{equation}
such that $\frac{dQ^{\pm}}{dt} =\{Q^{\pm},H \}+ \partial_t Q^{\pm}=0$. Those dynamical  variables take complex values admitting the polar decomposition $Q^{\pm}=q_0\, e^{\pm i \theta_0}$ and allowing in fact to determine the motion, which turns out to be bounded for $E=-|E| <0$. 
Indeed we have:
\begin{equation}
 \biggl(\mp i r p +  r \sqrt{2 m |E|}  -\sqrt{\frac{m}{2}}\frac{(k-\eta |E|)}{\sqrt{|E|}} \biggl)e^{\mp i \bigl(\sqrt{\frac{2}{m}}\frac{r p \sqrt{|E|}}{(k+\eta |E|)}+\sqrt{\frac{2}{m}}\,\frac{2 |E| \sqrt{|E|}}{(k+\eta |E|)}t \bigl)}=q_0e^{  \pm i \theta_0} \, ,
\label{moto}
\end{equation}
or else
\begin{equation}
\begin{cases}
- i r p +  r \sqrt{2 m |E|}  -\sqrt{\frac{m}{2}}\frac{(k-\eta |E|)}{\sqrt{|E|}} =q_0 \, e^{ i\bigl(\sqrt{\frac{2}{m}}\frac{r p \sqrt{|E|}}{(k+\eta |E|)}+\sqrt{\frac{2}{m}}\,\frac{2 |E| \sqrt{|E|}}{(k+\eta |E|)}t +\theta_0 \bigl)} \\
+ i r p +  r \sqrt{2 m |E|}  -\sqrt{\frac{m}{2}}\frac{(k-\eta |E|)}{\sqrt{|E|}} =q_0 \, e^{- i \bigl(\sqrt{\frac{2}{m}}\frac{r p \sqrt{|E|}}{(k+\eta |E|)}+\sqrt{\frac{2}{m}}\,\frac{2 |E| \sqrt{|E|}}{(k+\eta |E|)}t +\theta_0 \bigl)}\, ,
\end{cases}
\label{ecoupled}
\end{equation}
where $q_0=\sqrt{-l^2+\frac{m (k-\eta|E|)^2}{2 |E|}}$ (following from $A^+A^- +\gamma(H)=q_0^2+\gamma(H)=-l^2$). Summing and subtracting \eqref{ecoupled} we obtain:
\begin{equation}
\begin{cases}
 2 r \sqrt{2 m |E|}  - \sqrt{2 m}\frac{(k-\eta |E|)}{\sqrt{|E|}} = 2 q_0  \cos \biggl(\sqrt{\frac{2}{m}}\frac{r p \sqrt{|E|}}{(k+\eta |E|)}+\sqrt{\frac{2}{m}}\,\frac{2 |E| \sqrt{|E|}}{(k+\eta |E|)}t +\theta_0 \biggl) \\
 r p = - q_0  \sin \biggl(\sqrt{\frac{2}{m}}\frac{r p \sqrt{|E|}}{(k+\eta |E|)}+\sqrt{\frac{2}{m}}\,\frac{2 |E| \sqrt{|E|}}{(k+\eta |E|)}t +\theta_0 \biggl)\, .
\end{cases}
\label{eqcoupled}
\end{equation}
It is immediate to verify that taking the sum of the square of these two equations we obtain the equation $\eqref{factor}$ restricted to the level surface $H=-|E|$.
\noindent Finally, thanks to the above relations, we are able to obtain  $t$ as a  function of $r$:
\begin{equation}
\small{t(r)=\frac{1}{\Omega_{(\eta)}(E)}\biggl[\arccos\biggl(-\sqrt{\frac{m}{2}}\frac{\bigl((k-\eta |E|)-2|E|r\bigl)}{ q_0\sqrt{|E|}}\biggl)-\sqrt{\frac{2}{m}}\frac{\sqrt{|E|}}{(k+\eta|E|)}\sqrt{ 2 m r (k-\eta|E|)-2 m |E| r^2-l^2}-\theta_0 \biggl]}\, ,
\label{eq:traject}
\end{equation}
\noindent where $\Omega_{(\eta)}(E)=\sqrt{\frac{2}{m}}\,\frac{2 |E| \sqrt{|E|}}{(k+\eta |E|)} \equiv \alpha(E)$ is the angular frequency of the  motion.
\noindent Concerning (\ref {eq:traject}) it is evident that, due to the presence of  the ``inverse cosine'' function, $t$ is a multivalued function of $r$ defined mod $2\pi/\Omega$. To recover univaluedness, we have to introduce a ``uniformization map'' which is trivially given by the  periodic function $\cos (\Omega t)$.
In the limit $\eta \to 0$, the results  for the flat Kepler-Coulomb are recovered (see \cite {Kuru}). 
We can say that the motion has been \emph{algebraically determined}.

A number of plots are reported, showing the behavior of $V_{eff}(r) \doteq \frac{l^2}{2mr(r+\eta)}-\frac{k}{r+\eta}$ as a function of $r$, and the orbits on the phase plane $(r, p)$ for different values of the deformation parameter (for $2 l=m=k=1$, $E=-1$, in appropriate units). 

\begin{figure}[htbp]
\centering
\includegraphics[width=7cm, height=5cm]{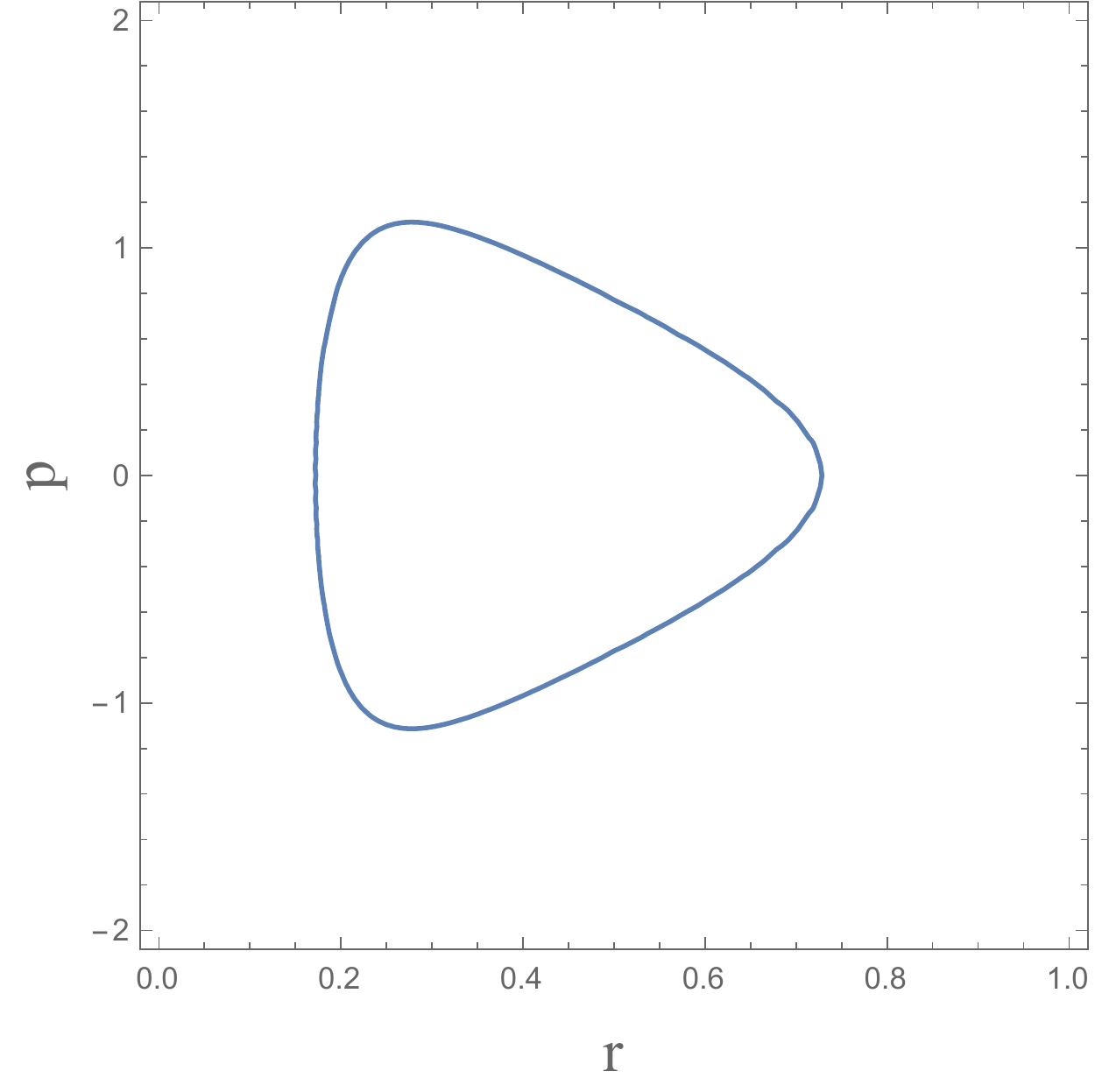} \quad 
\includegraphics[width=7cm, height=6cm]{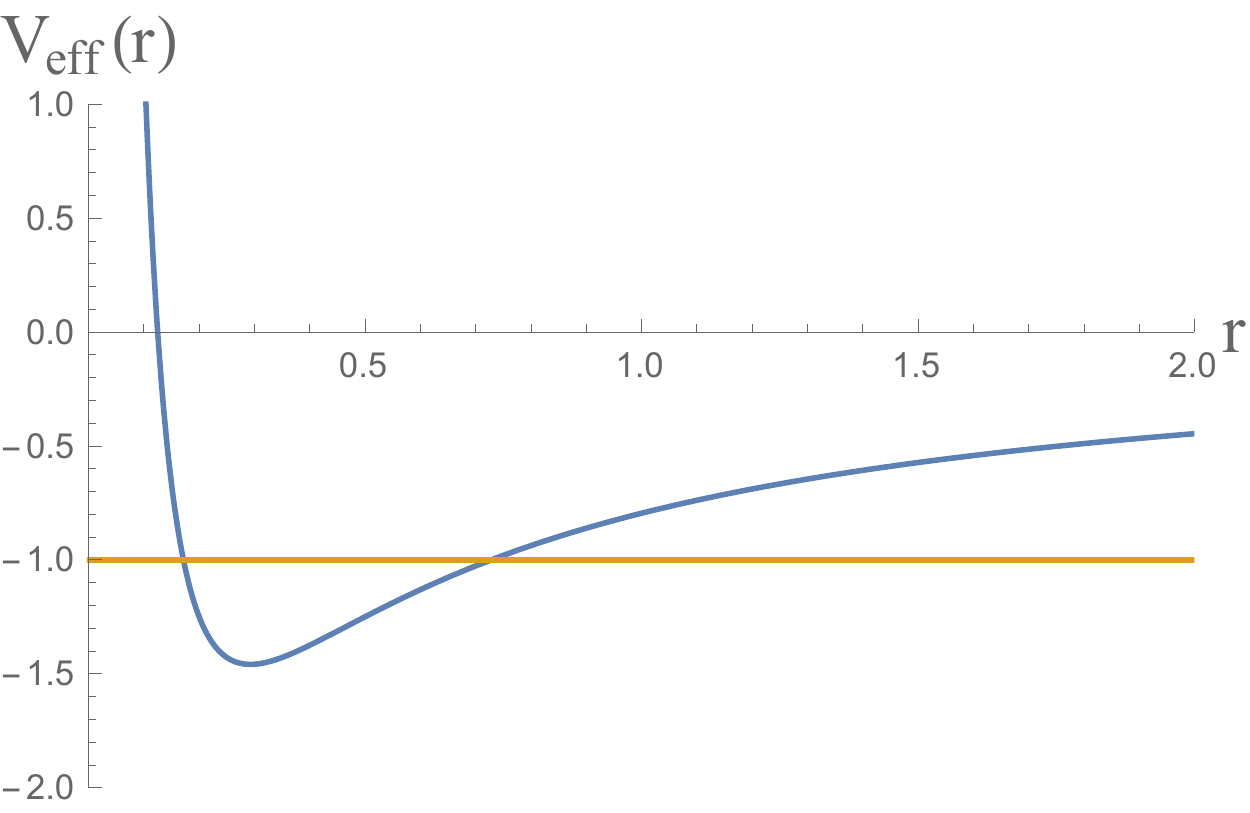}
\caption{Phase plane $(r, p)$ and effective potential $V_{eff}(r)$ for $\eta=0.1$ . }
\label{one}
\end{figure}
\begin{figure}[htbp]
\centering
\includegraphics[width=8cm, height=6cm]{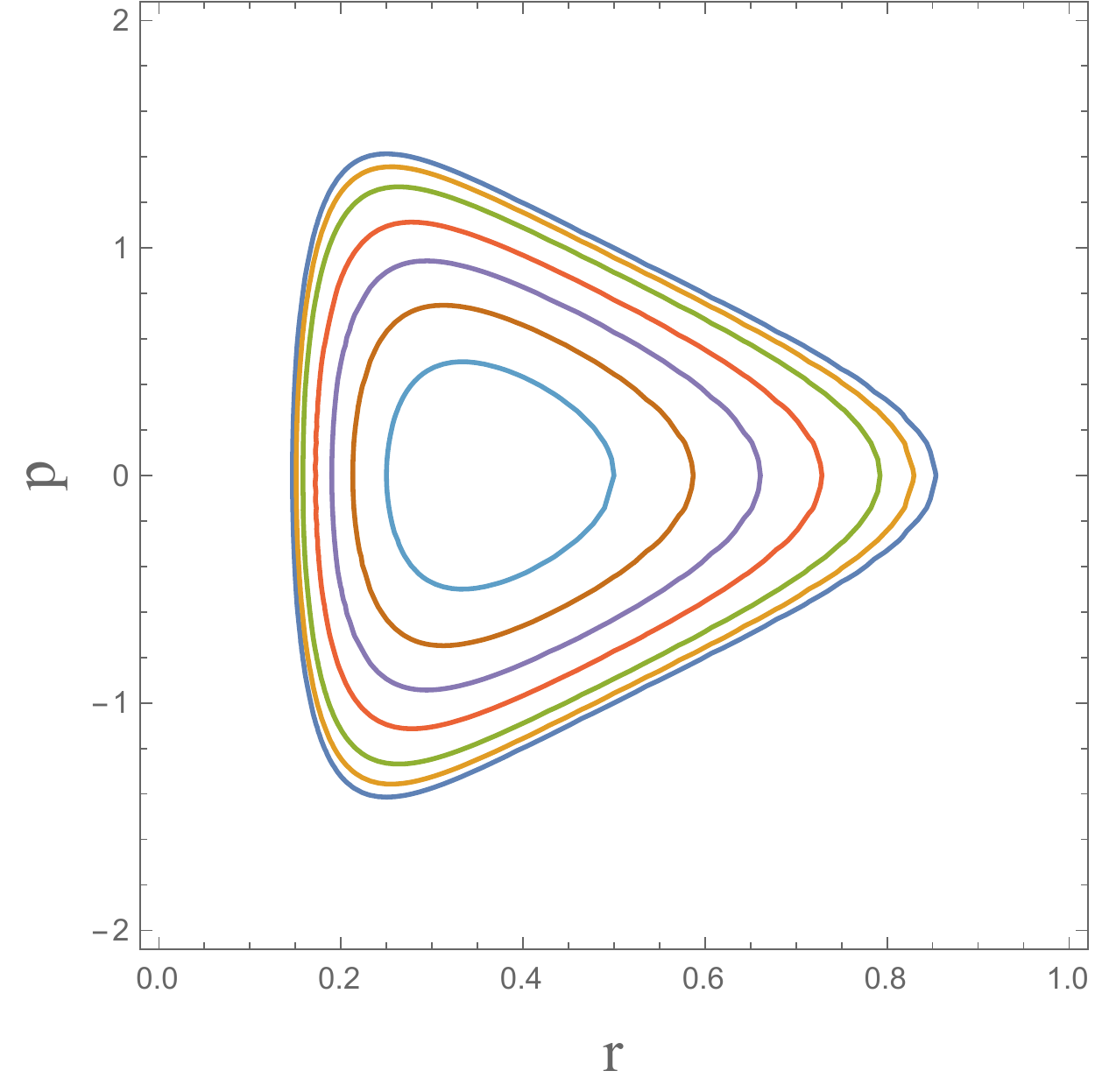} 
\caption{Phase plane $(r, p)$ for $\eta=0, 0.01, 0.05, 0.1, 0.15, 0.2, 0.25$ .}
\label{two}
\end{figure}

\section{Explicit formula for the orbits and third Kepler law}

\noindent  As we have shown in the introduction,   our system is maximally superintegrable and this maximal superintegrability is strictly related to the existence of the Runge-Lenz vector: then, as it happens for  the standard Kepler-Coulomb system,  we expect that this extra symmetry will play a crucial role in determining the shape of the orbits. As is well known, in the undeformed case the orbits are conic sections, namely ellipses for bounded trajectories.
\noindent To identify the analytic form of the orbits when $\eta \ne 0$, we will first consider the simplest and more \emph{physical} case, corresponding to  $\eta>0$. To  this end, we will closely follow Refs. \cite{Gold, CarGal}.

\noindent
 In  $\mathbb{R}^3$ the Runge-Lenz vector $\mathbf{R}$, when evaluated \emph {on-shell},  can be written as:
\begin{equation}
\mathbf{R} = \frac{1}{m}[\mathbf{p}(\mathbf{p} \cdot \mathbf{q})-\mathbf{q}(\mathbf{p}\cdot \mathbf{p})]+
\frac{\mathbf{q}}{|\mathbf{q}|}(k-\eta|E|) \, .
\label{runge}
\end{equation}
\noindent
Again, we see that its  expression   is formally identical to the one holding in the flat case and is obtained by letting $k\to k-\eta|E| :=K$.
\noindent
For its square we can write (again \emph{on-shell}):
\begin{equation}
\mathbf{R}^2 =  K^2 -\frac{2l^2|E|}{m}  \, ,
\label{value}
\end{equation}
\noindent and then
\begin{equation}
-\frac{2l^2}{m}\biggl(|E|-\frac{K}{r}\biggl) =\mathbf{R}^2 +K^2 +2 K |\mathbf{R}| \cos (\theta-\theta_0) \, .
\label{for}
\end{equation}
\noindent At this point, by elementary algebraic manipulations, it is easy to write the equation for the orbits in terms of $r:= |{\mathbf q}|$ and $\theta$, getting:
\begin{equation}
r(\theta)= \frac{\emph{p}_{(\eta)}}{1+ \epsilon_{(\eta)}\cos(\theta-\theta_0)} \, ,
\label{orbits}
\end{equation}
\noindent
($\emph{p}_{(\eta)}$ being the \emph{parameter} and $\epsilon_{(\eta)}$ the \emph{eccentricity} of the ellipses) which is formally the same expression holding in the flat case.
But now we have:
\begin{equation}
\begin{cases}
\emph{p}_{(\eta)} \equiv \emph{p}(E, \eta) = \frac{l^2}{m K}\\
\epsilon_{(\eta)}  \equiv \epsilon(E, \eta)=\frac{|\mathbf{R}|}{K} \, ,
\end{cases}
\label{eq}
\end{equation}
\noindent so that $\epsilon_{(\eta)}^2 = 1 - 2 |E| l^2/m K^2$.
\noindent In the above expression $\theta$ and $\theta_0$ are the angles that the vectors $\mathbf{q}$ and  $\mathbf{R}$
form with the half-line $\theta = 0$ (of course $\theta_0$ is a constant of the motion).

To check whether the third Kepler's law holds in the deformed case as well, we have to  compute  the ratio $\tau^2/a^3 = 4\pi^2/a^3\Omega^2$ where 
\begin{equation}
\Omega_{(\eta)}(E)=\frac{4|E| \sqrt{|E|}}{\sqrt{2m}(k+\eta|E|)} \, ,
\label{pulsazione}
\end{equation}
and $a$ is the larger semi-axis defined as $a =\frac{r_+ + r_-}{2}$. The inversion points $r_\pm$ 
(where $p_{r_+}=p_{r_-}=0$) are obtained by taking the roots of 
\begin{equation}
r^2 -\frac{(k-\eta |E|)}{|E|}r+\frac{l^2}{2m|E|}=0 \,\, \Rightarrow \, \, r_{(\eta)\pm}=\frac{k-\eta|E|}{2|E|} \pm \sqrt{\frac{(k-\eta|E|)^2}{4|E|^2}-\frac{l^2}{2m|E|}} \, ,
\label{eq:points}
\end{equation}
entailing
\begin{equation}
a_{(\eta)}=\frac{r_{(\eta)+}+r_{(\eta)-}}{2}=\frac{k-\eta|E|}{2|E|} \, .
\label{eq:kepler}
\end{equation}
In the limit $\eta \to 0$ we recover the larger semi-axis of the flat case, and then
\begin{equation}
\frac{\tau^2}{a^3}=\frac{4 \pi^2 m}{k} \, .
\label{eq:ris}
\end{equation}
We remind that the so-called  third Kepler's law is obtained by assuming that the ratio $\frac {m}{M}$  between the mass of the planet and the mass of the sun be very small, so that the reduced mass can be identified with the mass of the planet, entailing $k = GMm$ and thus
$\frac{\tau^2}{a^3}=\frac{4 \pi^2 m}{k}= \frac{4 \pi^2}{GM}$.
\noindent In the deformed case the analogous formula reads:
\begin{equation}
\frac{\tau_{(\eta)}^2}{a_{(\eta)}^3}=4 \pi^2 m\frac{(k+\eta|E|)^2}{(k-\eta|E|)^3} \, .
\label{etathird}
\end{equation}
\noindent
The Kepler's third law is then violated as the r.h.s of \eqref{etathird}, againassuming $k=GMm$, keeps its dependence upon $m$ and $E$:
\begin{equation}
\frac{\tau_{(\eta)}^2}{a_{(\eta)}^3}=4 \pi^2 m \frac{(k+\eta|E|)^2}{(k-\eta|E|)^3} =\frac{4 \pi^2}{GM}\bigl(1+
\frac{5\eta|E|}{GMm} + O(\eta^2)\bigl) \, .
\label{third}
\end{equation}
Some comments are in order, as the formulas we have derived seem to imply a sort of difference between the classical and the quantum case. Namely, according to the results obtained in \cite{Annals2013}
in the quantum case, for $\eta>0$ one has a very simple \emph{coupling constant metamorphosis}, amounting just to replace $k$ by $k+\eta E$. This substitution holds both for the spectrum and for the eigenfunctions.
We have already seen in \cite{PD} that a similar simple substitution applies for the quantum D-III as well. However in the \emph{classical}  Taub-Nut case, in order to close the Poisson algebra, one has to cope both with $k-\eta E$ \emph{and} with $k+\eta E$. An analogous  behaviour is exhibited by the classical D-III \cite{future}. 

\section{Explicit evaluation of the trajectory and comparison with algebraic method}

\noindent For the sake of completeness we present here the explicit derivation of the trajectory $t(r)$ using the standard analytic method \cite{Gold}. A comparison with the results obtained through the Spectrum Generating Algebra will provide a definite proof of the correctness of the algebraic approach. The starting point is the usual Hamiltonian \eqref{ham}:
\begin{equation}
H=\frac{r}{r+\eta}\biggl[\frac{p^2}{2m}+\frac{l^2}{2 m r^2}-\frac{k}{r}\biggl] \, .
\end{equation}
The radial momentum $p$ is related to the radial component of the velocity through the Hamilton's equation:
\begin{equation}
\dot r =\partial_p H=\frac{r}{r+\eta} \frac{p}{m}\,\, \Rightarrow \,\, p=\frac{r+\eta}{r} m \dot r \, .
\end{equation}
Inserting in the Hamiltonian the expression of $p$ in terms of $r$ and  $\dot r$ we obtain: 
\begin{equation}
H=\frac{r}{r+\eta}\biggl[\frac{(r+\eta)^2}{r^2}\frac{m}{2}\dot r^2+\frac{l^2}{2 m r^2}-\frac{k}{r}\biggl] \, .
\label{eq:Ham}
\end{equation}
By solving the above expression with respect  to $\dot r(t)$ and setting  $H=E$ we get:
\begin{equation}
\dot r(t) = \pm \sqrt{\frac{2}{m}} \frac{r}{r+\eta} \sqrt{E+\frac{k+\eta E}{r}-\frac{l^2}{2 m r^2}} \, .
\label{eq:velocity}
\end{equation}
Comparing with the Euclidean case ($\eta = 0$),  besides the coupling constant metamorphosis, the essential difference consists in  the presence of a nontrivial conformal factor.  As a next step, we calculate $t(r)$ by taking the positive branch of the square root: 
\begin{equation}
\small{t(r)-t_0= \sqrt{\frac{m}{2}}  \int_{r_0}^r dr \frac{r+\eta}{r \sqrt{E+\frac{k+\eta E}{r}-\frac{l^2}{2 m r^2}}}=\sqrt{\frac{m}{2}}  \int_{r_0}^r \frac{dr}{ \sqrt{E+\frac{k+\eta E}{r}-\frac{l^2}{2 m r^2}}}+\sqrt{\frac{m}{2}}\eta \int_{r_0}^r \frac{dr}{r \sqrt{E+\frac{k+\eta E}{r}-\frac{l^2}{2 m r^2}}}} \, .
\label{eq:tra}
\end{equation}
The two integrals involved in the above formula can be conveniently calculated by introducing the so-called \emph{eccentric anomaly} $\Psi_{(\eta)}$ through the relation \cite{Gold}:
\begin{equation}
r=a_{(\eta)}(1-\epsilon_{(\eta)}\cos \Psi_{(\eta)}) \, .
\end{equation}
In the previous section we have already shown that the semi-major axis is given by 
$a_{(\eta)}=-\frac{k+\eta E}{2 E}$ and the eccentricity reads $\epsilon_{(\eta)}=\sqrt{1+\frac{2l^2E}{m(k+\eta E)^2}}$. 
Let us  now pass to the explicit calculation of the  two integrals contained in (\ref{eq:tra}), setting there $E=-|E|<0$.
It is not too difficult to arrive at the following results:
\begin{align}
 \sqrt{\frac{m}{2}}  \int_{r_0}^r \frac{dr}{\sqrt{-|E|+\frac{k-\eta |E|}{r}-\frac{l^2}{2 m r^2}}}&=\sqrt{\frac{m a_{(\eta)}^3}{k-\eta|E|}}\int_0^{\Psi_{(\eta)}} d\Psi'_{(\eta)} (1-\epsilon_{(\eta)}\cos \Psi'_{(\eta)}) \nonumber \\
&=\sqrt{\frac{m a_{(\eta)}^3}{k-\eta|E|}} (\Psi_{(\eta)} -\epsilon_{(\eta)} \sin \Psi_{(\eta)}) \, ,
\end{align}
\begin{equation}
\sqrt{\frac{m}{2}}\eta \int_{r_0}^r \frac{dr}{r \sqrt{-|E|+\frac{k-\eta |E|}{r}-\frac{l^2}{2 m r^2}}}=\sqrt{\frac{m a_{(\eta)}}{k-\eta|E|}} \eta \int_0^{\Psi_{(\eta)}} d\Psi'_{(\eta)}=\sqrt{\frac{m a_{(\eta)}} {k-\eta|E|}} \eta \Psi_{(\eta)} \, .
\label{eq:int}
\end{equation}
Hence, dividing and multiplying  the output of the second integral by the same quantity $a_{(\eta)}$ and rearranging the two integrals  in a single  expression, we get the trajectory  (with the initial condition $t_0=0$):
\begin{equation}
t(r)= \sqrt{\frac{m a_{(\eta)}^3} {k-\eta|E|}} \biggl(\frac{\eta+a_{(\eta)}}{a_{(\eta)}}\biggl)\Psi_{(\eta)}-\sqrt{\frac{m a_{(\eta)}^3} {k-\eta|E|}}\epsilon_{(\eta)}\sin \Psi_{(\eta)} \, ,
\label{eq:keep}
\end{equation}
namely:
\begin{align}
t(r)&= \sqrt{\frac{m a_{(\eta)}^3} {k-\eta|E|}} \biggl(\frac{\eta+a_{(\eta)}}{a_{(\eta)}}\biggl)\biggl[\Psi_{(\eta)}-\frac{ a_{(\eta)}}{\eta +a_{(\eta)}}\epsilon_{(\eta)}\sin \Psi_{(\eta)}\bigg] \nonumber \\
&=\frac{1}{\Omega_{(\eta)}(E)}\biggl[\Psi_{(\eta)}-\frac{ a_{(\eta)}}{\eta +a_{(\eta)}}\epsilon_{(\eta)}\sin \Psi_{(\eta)}\bigg] \, ,
\label{eq:kepl}
\end{align}
which is \emph{the deformed  Kepler equation}:
\begin{equation}
\Omega_{(\eta)}(E) t(r)=\Psi_{(\eta)}-\frac{ a_{(\eta)}}{\eta +a_{(\eta)}}\epsilon_{(\eta)}\sin \Psi_{(\eta)} \, .
\label{eq:kepler}
\end{equation}
The frequency of the motion is given by
\begin{equation}
\Omega_{(\eta)}(E)=\sqrt{\frac{k-\eta |E|}{m a^3_{(\eta)}}}\frac{a_{(\eta)}}{\eta+a_{(\eta)}}=\sqrt{\frac{2}{m}} \frac{2|E|\sqrt{|E|}}{k+\eta|E|} \, ,
\label{pulsazione}
\end{equation}
which is nothing but the same frequency obtained  through the Spectrum Generating Algebra. Now we have just to plug in the equation (\ref{eq:kepler}) the explicit form of $\Psi_{(\eta)}$ and check whether it coincides with the one derived via the algebraic method. By solving for $\Psi_{(\eta)}$ one gets  

\begin{equation}
\Psi_{(\eta)}=\arccos \biggl[\frac{1}{\epsilon_{(\eta)}}\biggl(1-\frac{r}{a_{(\eta)}}\biggl)\biggl] \, ,
\label{psi}
\end{equation}
whence, owing to the  well known relation $\sin(\arccos(x))=\sqrt{1-x^2}$, it follows
\begin{equation}
\Omega_{(\eta)}(E) t(r) = \arccos \biggl[\frac{1}{\epsilon_{(\eta)}}\biggl(1-\frac{r}{a_{(\eta)}}\biggl)\biggl] -\frac{a_{(\eta)}}{\eta +a_{(\eta)}} \sqrt{\epsilon_{(\eta)}^2-\biggl(1-\frac{r}{a_{(\eta)}}\biggl)^2} \, .
\label{ke1}
\end{equation}
Equation (\ref{ke1}) represents the trajectory calculated  through the  standard analytic method. 

\noindent On the other hand, the equation \eqref{eq:traject} for  the trajectory derived by means of the algebraic method yields (in the case $\theta_0=0$): 
\begin{equation}
\small{\Omega_{(\eta)}(E)t(r)=\arccos\biggl(-\sqrt{\frac{m}{2}}\frac{\bigl((k-\eta |E|)-2|E|r\bigl)}{ q_0\sqrt{|E|}}\biggl)-\sqrt{\frac{2}{m}}\frac{\sqrt{|E|}}{(k+\eta|E|)}\sqrt{ 2 m r (k-\eta|E|)-2 m |E| r^2-l^2}}\, ,
\label{eq:trajects}
\end{equation}
where $q_0=\sqrt{-l^2+\frac{m(k-\eta|E|)^2}{2|E|}}$. After easy algebraic manipulations equation \eqref{eq:trajects}
acquires the form:
\begin{equation}
\Omega_{(\eta)}(E) t(r) = \arccos \biggl[-\frac{1}{\epsilon_{(\eta)}}\biggl(1-\frac{r}{a_{(\eta)}}\biggl)\biggl] -\frac{a_{(\eta)}}{\eta +a_{(\eta)}} \sqrt{\epsilon_{(\eta)}^2-\biggl(1-\frac{r}{a_{(\eta)}}\biggl)^2} \, .
\label{ke}
\end{equation} 
In other words, by the algebraic method we get  $t(r)$ evaluated for $-\epsilon_{(\eta)}$. As we expected, this result is just  the  $\eta-$deformation of the Kuru-Negro result \cite {Kuru}. 
\vskip 2pt
\section{The case $\eta<0$ : new features}
\noindent This section is  devoted to a terse investigation of the main features arising in the case  $\eta<0$. In this case the conformal factor $\frac{r}{r+\eta}$ can be more conveniently written as $\frac{r}{r-|\eta|}$ which \emph{emphasizes} the singularity at $r=|\eta|$.
One relevant question is whether the singularity can be overcome or not. In the first case there might be trajectories intersecting the line $r=|\eta|$. In the second case the phase plane ($r$, $\dot r$) will consist of two non overlapping domains. In particular, for closed orbits one may ask under what conditions the following (mutually excluding) inequalities for the inversion points hold:
\begin{equation}
\label{rminmax}
r_{(\eta)-} > |\eta| \, ,
\, r_{(\eta)+}<|\eta| \, .
\end{equation}

\noindent
A careful analysis of \eqref{rminmax} shows that to characterise the corresponding regions of this plane one has to  look at  \emph{both} parameters $\eta$ and  $\lambda$, a characteristic lenght scale defined as  
$\lambda:=\frac{l^2}{2mk}$, or better at their ratio $\alpha := \frac{|\eta|}{\lambda} $,  and at the behaviour of the effective potential $V_{eff}(r)=\frac{l^2}{2 m r (r-|\eta|)}-\frac{k}{r-|\eta|}=-\frac{k}{r}[\frac{r-\lambda}{r-\alpha \lambda}]$.  
\subsection{case $\alpha<1$}

\noindent The most interesting situation occurs in  the case $\alpha <1$, where one has indeed  two  non-overlapping regions separated by  the straight-line $r = |\eta|$.

\begin{figure}[htbp]
\centering
\includegraphics[width=9cm, height=6cm]{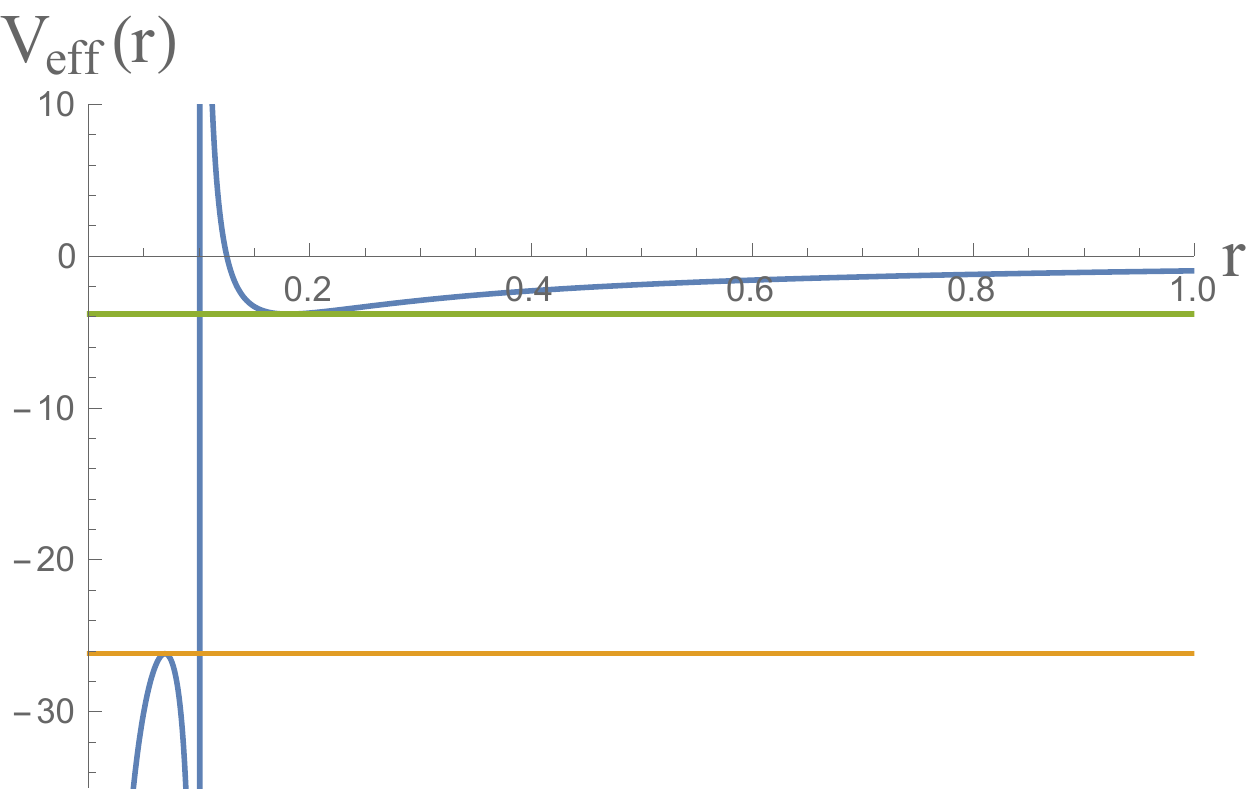} \quad 
\caption{Potential $V_{eff}(r)$ for $\alpha=\frac{4}{5}$. The straight lines represent the Energies associated to the critical points.}
\label{three}
\end{figure}
\begin{itemize}
\item In the right domain $r>|\eta|$ the conformal factor is positive.  We have a Riemannian manifold with non constant curvature and there will be closed trajectories whenever the energy belongs to the (negative) open interval $\bigl(0, V_{eff} (r_+)\bigl)$, where $V_{eff}(r_+)=\frac{- k}{\lambda (1 +\sqrt{1 -\alpha})^2}$ is the value of the effective potential at the critical point $r_+ = \lambda (1 +\sqrt{1 -\alpha})$.
\item In the left domain $r<|\eta|$ the conformal factor is \emph{negative} entailing that the kinetic energy is also  negative. In order to get a physically significant system we are naturally led to define in this region a new Hamiltonian $\widetilde{\mathcal{H}}:= -\mathcal{H}=\frac{r}{|\eta|-r}\frac{p^2}{2 m}+\widetilde{V}_{eff}(r)$ with $\widetilde{V}_{eff}(r):= \frac{l^2}{2 m r (|\eta|-r) }-\frac{k}{|\eta|-r}$, namely to look at the  system obtained by \emph{time-reversal}.
As it is clearly shown by \figref{four} after that transformation in  the region $0<r<|\eta|$ \emph{the effective potential acquires a typical  "confining" shape}.  There will be closed orbits for any positive energy higher than $\widetilde{V}_{eff}(r_-)$, where $r_-= \lambda (1 -\sqrt{1 -\alpha})$. We point out that the minimum of the potential is a monotonically decreasing function of $|\eta|$, so that it goes to infinity as $|\eta|$ goes to zero.
\end{itemize}
\begin{figure}[htbp]
\centering
\includegraphics[width=9cm, height=6cm]{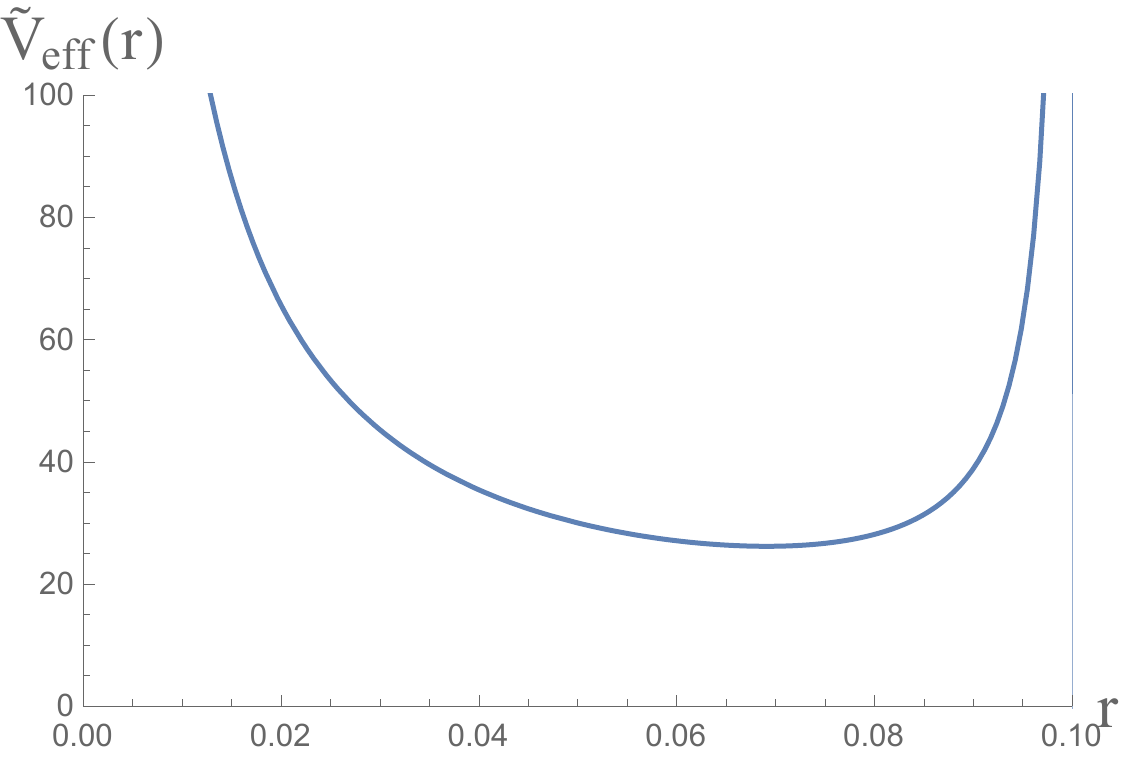} \quad 
\caption{Potential $V_{eff}(r)$ after the time-reversal transformation, i.e. $\widetilde{V}_{eff}(r):=-V_{eff}(r)$ calculated for $\alpha=\frac{4}{5}$. The latter is contained into the segment $0<r<|\eta|$.}
\label{four}
\end{figure}

\subsection{case $\alpha>1$}
\noindent  Here the dynamics is certainly less interesting because no closed orbits will come out. 
\begin{figure}[htbp]
\centering
\includegraphics[width=9cm, height=6cm]{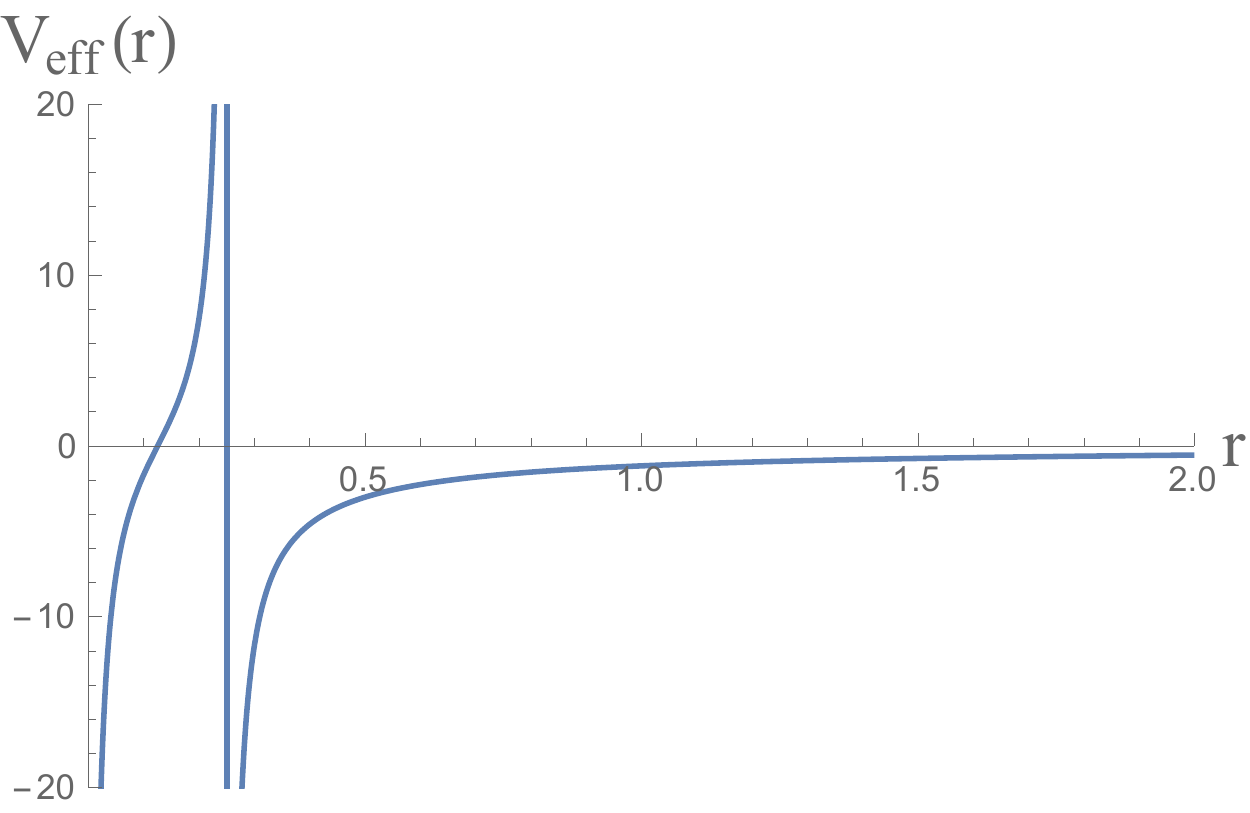} \quad 
\caption{Potential $V_{eff}(r)$ calculated for $\alpha=2$.}
\label{five}
\end{figure}

\noindent In the region $r>|\eta|$ the effective potential will be proportional to $-(r-|\eta|)^{-1}$ while in the bounded region $0<r<|\eta|$ its image is the full real line and furthermore it exhibits an inflection point for the value $\bar r= \lambda\bigl(1-(\alpha-1)^{1/3}+(\alpha-1)^{2/3}\bigl)$.
\subsection{case $\alpha=1$} 
\noindent On the boundary line $\alpha=1$ things are definitely less clear. In fact, it looks like that the singularity could be overcome.  By the way, a plot of the effective potential for $\alpha=1$, i.e. $V_{eff}(r)=-\frac{k}{r}$, shows that the distinction between the two regions ($r>|\eta|$, $r<|\eta|$) disappears, in the sense that we have a single continuous line  with a monotonically increasing behaviour and of course no closed orbits are allowed. However, at the same time \eqref{eq:velocity} implies that at $r=|\eta|$ the (absolute value of) the velocity diverges.

\begin{figure}[htbp]
\centering
\includegraphics[width=9cm, height=6cm]{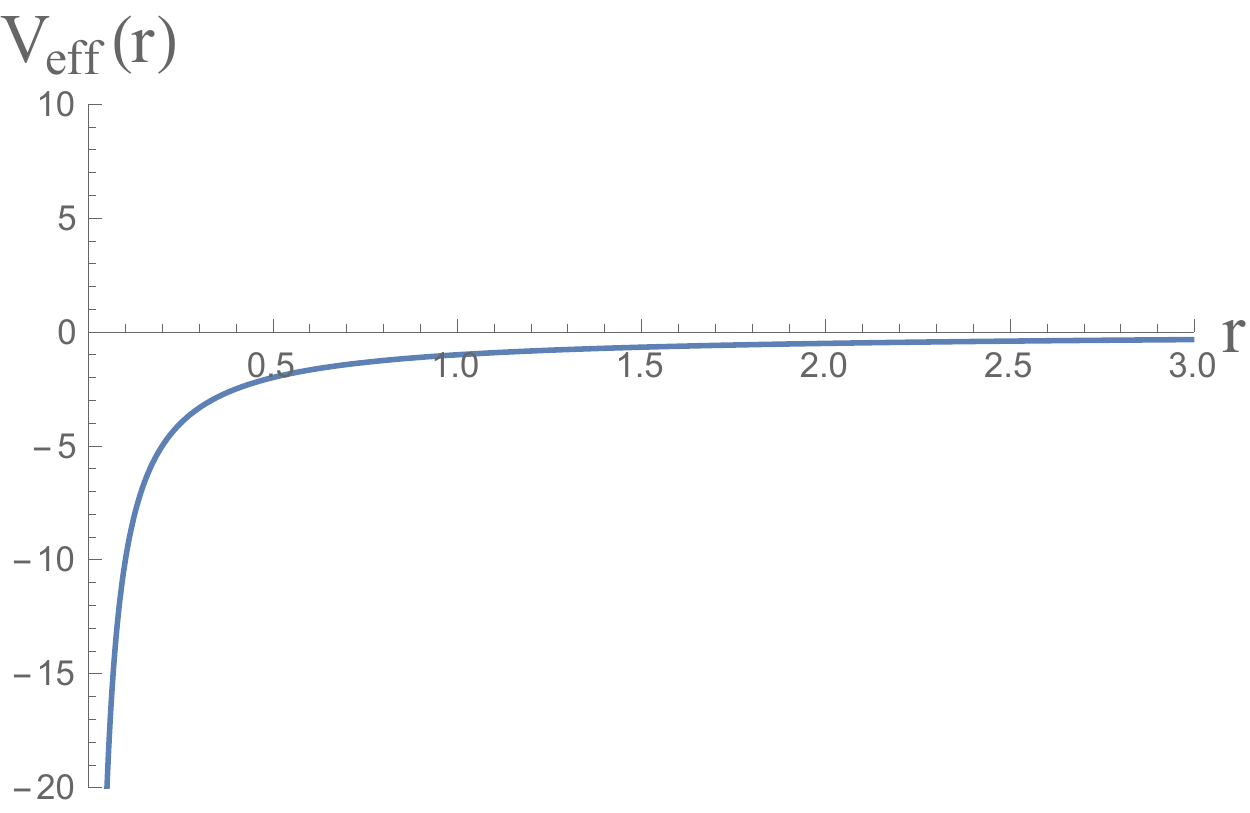} \quad 
\caption{Potential $V_{eff}(r)$ calculated for $\alpha=1$. The centrifugal and gravitational contributions add up to gives a behaviour equals to $- k r^{-1}$. In this case the singularity in the effective potential disappears.}
\label{five}
\end{figure}

\section{Concluding Remarks and Open Problems}

\noindent
One of the main results obtained in our paper is the constructive proof that the Spectrum Generating Algebra technique can be successfully employed to attack and solve maximally superintegrable systems \emph{on spaces with variable curvature}. As already mentioned throughout the article, in the next future we will provide the analogous results for the classical D-III system. Moreover, for \emph{positive values} of the deformation parameters, we will exhibit the exact solution to the corresponding quantum problems based on the \emph{Shape Invariant Potentials} techniques \cite{future}, making a comparison with different approaches proposed in the literature \cite{Annals2013, 2014arXiv1410.4495P}. 

As a further interesting result, it is worth to stress that we have shown the existence of closed orbits even for negative values of the deformation parameter. As a matter of fact, the behaviour of the classical effective potential  strongly suggests that  in the quantum case there will be bound states also in the region $0<r<|\eta|$, while in the limit $|\eta| \to 0$ the minimum of the potential will go to infinity. With the proper changes, we expect that similar features will hold for D-III as well.

Actually, in this context our final aim is twofold:

\begin{enumerate} 

\item  we will focus future investigations on the quantum systems exactly on the case where the deformation parameters take negative values. There, due to the confining nature of the potentials, we expect the most interesting results from a physical point of view.

\item we will try to solve all the classical problems belonging to the Perlick's families I and II \cite{Perlick} by means of the Spectrum Generating Algebra approach. We are encouraged to proceed further in this direction inasmuch as we have seen that  in the classical deformed versions of Taub-Nut and D-III the \emph{coupling constant metamorphosis}, hardly applicable to the full Perlick's families, does not seem to be  \emph{the} essential feature. 

\end{enumerate}

\vskip 2cm

%%%%%%%%%%%%%%%%%%%%%%%%%%%%%%%%%%%%%%%%%%%%%%%%%%%
\subsection*{Acknowledgments}

\noindent This work was partially supported by the grant AIC-D-$2011$-$0711$ (MINECO-INFN)  (O. R.), by the italian MIUR under the project PRIN $2010$-$11$ (Analytical and geometrical aspects of finite and infinite-dimensional hamiltonian systems, prot. n. $2010$JJ$4$KPA\_$004$). The authors acknowledge with pleasure enlightening discussions with the spanish and italian colleagues  A. Ballesteros and F. J. Herranz  (Departamento de Fisica, UBU, Spain), D. Riglioni (CRM, Montreal, Canada), F. Zullo (Math-Phys. Dept. Roma Tre, Italy) .

\end{document}